\documentclass[aps,prl,reprint,nofootinbib]{revtex4-2}

\usepackage{graphicx}% Include figure files
\usepackage{dcolumn}% Align table columns on decimal point
\usepackage{bm}% bold math
\usepackage{hyperref}% add hypertext capabilities
\usepackage{latexsym,amssymb,amstext,amsmath}

\usepackage{xcolor}
\usepackage{mathtools}

\begin{document}

\preprint{Non-perpetual Eternal Inflation and the Emergent de Sitter Swampland Conjecture}

\title{Non-perpetual Eternal Inflation and the Emergent de Sitter Swampland Conjecture}% Force line breaks with \\

 %email{omerguleryuz@itu.edu..tr}
%\affiliation{%Authors' institution and/or address\\This line break forced with %\textbackslash\textbackslash}%


\author{Omer Guleryuz}
\homepage{omerguleryuz@itu.edu.tr}
\affiliation{
 Department of Physics, Istanbul Technical University, Maslak 34469 Istanbul, Türkiye%\\
}%
%\affiliation{Third institution, the second for Charlie Author
%\affiliation{% Authors' institution and/or address\\This line break forced with \textbackslash\textbackslash}%


\date{\today}%

\begin{abstract}
We introduce a novel correlation, $n_{\textrm{s}}$ - $\Delta N$, connecting CMB parameters to the required total e-folds for eternal inflation. This correlation provides a robust tool for evaluating eternal (string) inflation models using CMB data and explores the impact of quantum fluctuations on non-attractor phases. By generalizing eternal inflation criteria, our parameterization simplifies rigorous testing of predictions and reveals a link between refined de Sitter conjecture parameters and the eternal nature of the cosmic landscape. This points to a general tendency towards eternal behavior in low-energy effective field theories within the landscape, opening the possibility for our cosmic stage to potentially embrace a 'multiverse' scenario.
\end{abstract}

\maketitle
%\tableofcontents
\section{\label{sec:introduction} Introduction.} 
\vspace{-4mm}
In our exploration of cosmic understanding, the traditional notion of a solitary universe \cite{Starobinsky:1980te,Guth:1980zm,Linde:1981mu,Sato:1980yn,Albrecht:1982wi,SATO198166} has evolved into a nuanced investigation. We now focus on understanding realms where mini-universes, akin to ours, may exist. Eternal inflation, proposed by Linde \cite{Linde:1986fc,Linde:1986fd} and guided by anthropic principles, reveals a tapestry of universes forming a fractal pattern within the vast landscape of string theory. It offers a solution to the fine-tuning conundrum of the cosmological constant.

The universe, driven by an inflaton field, undergoes subatomic fluctuations, leading to variations in the amplitude and trajectory of the scalar field, resulting in eternal inflation. These quantum fluctuations generate primordial density fluctuations, leaving signatures on the cosmic microwave background (CMB) and large-scale structure. The crucial parameter in this cosmic history is the total number of e-foldings of inflation—a measure that extends beyond the observational reach of our scales yet remains fundamental.

In this letter\footnote{We employ natural units where $M_{\textrm{pl}} \equiv 1$.}, we present a simple generalization of eternal inflation criteria applicable to virtually every eternal inflationary model. This involves parameterizing a bound on the amplitude of curvature perturbation. We demonstrate how different eternal inflation phases' predictions can be explicitly tested against observational CMB data \cite{Planck:2018jri,Planck:2018vyg}, particularly regarding the primordial tilt and the total number of e-folds. Our parameterization establishes a meaningful link between the refined de Sitter Conjecture \cite{Ooguri:2018wrx,Garg:2018reu} and its emergence from eternal inflation criteria.

\vspace{-3mm}
\section{\label{sec:eternalinf}Non-perpetual eternal inflation.}
\vspace{-4mm}
Eternal inflation occurs when quantum jumps of a field $\phi$ dominate its classical rolling within a cosmic heartbeat, $H^{-1}$. Each domain of size $\sim H^{-1}$ perpetually gives rise to new domains, initiating instances of slow-roll (SR) evolution. This endless cosmic reset evolves for fields meeting the condition \cite{Starobinsky:1979ty,Mukhanov:1981xt,2003astro.ph..3077M,Linde:1983gd,Hawking:1982cz,Guth:1982ec,Starobinsky:1982ee,PhysRevD.28.679,Goncharov:1987ir,Guth:2007ng}:
\begin{equation}\label{eq1}
\frac{\delta \phi_{\textrm{qu}}}{\delta \phi_{\textrm{cl}}} = \frac{H^2}{2 \pi \Dot{\phi}} = P_{\mathcal{R}}(k) > 1,
\end{equation}
where
\begin{equation}\label{fluctuations}
\delta \phi_{\textrm{qu}}=\frac{H}{2\pi}\sqrt{H t}, \quad \delta \phi_{\textrm{cl}} =\Dot{\phi}t
\end{equation}
represent typical quantum and classical steps for the Hubble time scale, $t=H^{-1}$. $P_{\mathcal{R}}(k)$ denotes the curvature perturbation amplitude for horizon-crossing modes, given by { the ansatz near a reference scale $k_{*}$}\footnote{In all the examples considered throughout the paper, the assumption of a very mild scale-dependence (i.e., $n_{\textrm{s}} \simeq$ constant) is utilized, taking into account the observed CMB values.}:
\begin{equation}\label{eq2}
    P_{\mathcal{R}}(k)=P_{*} \left(\frac{k}{k_*}\right)^{n_{\textrm{s}}(k_*) -1}
\end{equation}
with $P_{\mathcal{R}}(k_*)=P_* \simeq 2.1 \times 10^{-9}$ and $k_* = 0.05 $ $\textrm{Mp{c}}^{-1}$ from CMB observations \cite{Planck:2018jri,Planck:2018vyg} { where the subscript '${}_*$' denoting the pivot scale.} For the eternal inflation condition $P_{\mathcal{R}}(k) > 1$, adopting the shallowest limit with $n_{\textrm{s}} < 0.98$ implies a requisite total e-folds number of $\Delta N \gtrsim 1000$ \cite{Kinney:2014jya}. Note that this derivation assumes a perpetual eternal inflation phase, as a deviation to $n_{\textrm{s}} < 0.94$ would yield $\Delta N \gtrsim 400$ { where the definition of the number of e-folds from the time a mode $k$ crosses the horizon until the end of inflation is given by:
\begin{equation}
     \Delta N \equiv N(k) = \ln{\frac{a_{\textrm{end}}}{a_k}},
\end{equation}
and it is assumed $a_{\textrm{k}}/a_{*} \simeq k/k_{*}$ under the approximation of $H \simeq H_{*}$ during inflation.

Following that, one can separate the total number of e-folds into two parts by considering the period before reaching the CMB observational scales (probing approximately $7$ e-folds around the pivot scale) and the evolution from the pivot scale to the end of inflation explicitly as:
\begin{equation}
\Delta N = \ln{\frac{a_{\textrm{end}}}{a_*}} + \ln{\frac{a_{*}}{a_k}} = N_{*} + \bar{N}.
\end{equation}
Here, the first term $N_{*}$ is the number of e-folds from the time the pivot scale crosses the horizon to the end of inflation ($k\geq k_{*}$) and usually takes values around $50-60$ e-folds to explain the current data \cite{Liddle:2003as}. The second term, $\Bar{N}$, is of interest in the eternal inflation era, assuming it takes place long before the observational reach of our scales ($k \ll k_{*}$). Since a successful solution of the horizon problem requires $\Delta N > N_{*}$, it follows that $\Bar{N} \neq 0$. 

When $\delta \phi_{\textrm{cl}} < |\delta \phi_{\textrm{qu}}|$, the inflaton field, responsible for driving inflation, can move both backward and forward with equal ease, swayed by the random whims of quantum fluctuations. If it moves backward, the return journey allows the universe to expand exponentially, creating new patches with similar quantum jumps. This scenario, where some regions remain in a state of never-ending inflation, is referred to as `perpetual eternal inflation.' Each exiting patch spawns countless others before it departs, perpetuating the inflationary process on a grand cosmic scale.

Yet, even in this seemingly endless inflation, there is an underlying certainty. The potential landscape exerts a pull towards its minimum, ensuring that every point in space will eventually find its way out of the inflationary phase. Thus, while inflation ends everywhere, the duration varies across different points, i.e. different $\Bar{N}$ values. Consequently, the ansatz (\ref{eq2}) is quantitatively valid for assuming a case where the curvature perturbation is continuously reducing until enough e-folds have occurred to reach the pivot scale, thereby determining the minimum necessary total number of e-folds. Thus,} examining eq. (\ref{eq2}) yields an alternative insight when solving for $n_{\textrm{s}}$:
\begin{equation}\label{eq3}
n_{\textrm{s}}{(k_{*})} = 1 - \frac{\ln{\left(\frac{P_{\mathcal{R}}}{P_{*}}\right)}}{\Delta N_{\textrm{min}}}.
\end{equation}
Considering constraints from CMB observational limits, denoted as $n_{\textrm{s,max}}{(k_{*})} > n_{\textrm{s}}{(k_{*})} > n_{\textrm{s,min}}{(k_{*})}$, we establish upper and lower bounds on the total e-folds number:
\begin{equation}\label{eq4}
\frac{\ln{\left(\frac{P_{*}}{P_{\mathcal{R}}}\right)}}{n_{\textrm{s,max}} -1} > \Delta N_{\textrm{min}} > \frac{\ln{\left(\frac{P_{*}}{P_{\mathcal{R}}}\right)}}{n_{\textrm{s,min}} -1}.
\end{equation}
{From a} local perspective, adhering to the shallowest curvature perturbations limit for an eternal inflation phase with a finite lifetime ($P_{\mathcal{R}} \approx 1$) and considering the observational value ($68\%,$ Planck TT,TE,EE+lowE+lensing) \cite{Planck:2018jri,Planck:2018vyg}, $n_{\textrm{s}} = 0.9649 \pm 0.0042$, the derived boundaries allow for the values of the total number of e-folds:
\begin{equation}
508.43 \lesssim \Delta N_{\textrm{min}} \lesssim 646.64.
\end{equation}

We refer to this stage as `non-perpetual eternal inflation,' characterized by a finite lifetime due to the curvature perturbation reaching its shallowest limit. Influenced by a tachyonic instability \cite{Kinney:2018kew}, this finite duration is perceived locally as non-perpetual, indicating a constrained existence for eternal inflation.

Similarly, quantum diffusion typically refers to the spread or dispersion of quantum fields. At its essence, it encapsulates the stochastic evolution of scalar fields during inflation. Since the quantum perturbations $\delta \phi_{\textrm{qu}}$ are stretched during inflation, they effectively appear as a nearly homogeneous classical field with a magnitude \cite{Linde:2007jn}:
\begin{equation}\label{eq21}
\Bar{\phi} \sim \delta \phi_{\textrm{qu}} = \frac{H}{2\pi}\sqrt{H t} \quad \rightarrow \quad \Dot{\delta \phi_{\textrm{qu}}} = \frac{H^3}{8\pi^2 \Bar{\phi}}.
\end{equation}
The condition for quantum diffusion with a typical time interval $t=H^{-1}$ is expressed as
\begin{equation}\label{eq22}
\frac{\Dot{\delta \phi_{\textrm{qu}}}}{|\Dot{\phi}|} = \frac{H^2}{4 \pi \Dot{\phi}} = \frac{P_{\mathcal{R}}(k)}{2} > 1.
\end{equation}
Compared with the eternal inflation condition, the bound on the amplitude of the curvature perturbation becomes twice as strong as $P_{\mathcal{R}}(k)>2.$ Consequently, using eq. (\ref{eq4}) when the quantum diffusion speed matches the classical speed ($P_{\mathcal{R}} \approx 2$), the allowed region for the total number of e-folds shifts to higher values.

\vspace{-3mm}
\section{\label{sec:stochastic}Stochastic approach.}
\vspace{-4mm}
A stochastic perspective on eternal inflation arises from examining classical inflation speeds, given by:
\begin{equation}\label{eq5}
\frac{\Dot{\phi}^2}{H^4} = \frac{3}{2 \pi^2}.
\end{equation}
This reveals a critical value leading to a sharp transition in the probability distribution for reheating volume \cite{Creminelli:2008es}. Sub-critical inflation speeds result in infinite distribution moments, yielding a non-vanishing probability of strictly infinite reheating volume alongside a finite probability for finite values. Thus, eternal inflation occurs if and only if \cite{Creminelli:2008es}:
\begin{equation}\label{eq6}
\frac{2 \pi^2 \Dot{\phi}^2}{3 H^4} < 1,
\end{equation}
which can be reorganized as $P_{\mathcal{R}} > 1/\sqrt{6}$. Interpreting this limit, one can rescale quantum fluctuations accordingly \cite{Hertog:2015zwh}:
\begin{equation}\label{eq7}
\delta \phi_{\textrm{qu}}^2 = \frac{C H^2}{2 \pi^2} \quad \overset{C=3}{\Longrightarrow} \quad \delta \phi_{\textrm{qu}} = \sqrt{\frac{3}{2}} \cdot \frac{H}{\pi},
\end{equation}
with fixed classical fluctuations, $\delta \phi_{\textrm{cl}} = \Dot{\phi} / H$. However, considering eq. (\ref{fluctuations}), this requires changing the time interval for quantum steps to $t=2CH^{-1}$ while keeping the time interval fixed for classical steps at $t=H^{-1}.$

Without re-scaling quantum fluctuations but conserving the quantum-to-classical ratio, the eternal inflation condition for $t=H^{-1}$ is expressed more generally as:
\begin{equation}\label{eq8}
\sqrt{2C} \cdot \frac{\delta \phi_{\textrm{qu}}}{\delta \phi_{\textrm{cl}}} = \sqrt{2C} P_{\mathcal{R}} > 1 \quad \rightarrow \quad P_{\mathcal{R}} > \frac{1}{\sqrt{2C}}
\end{equation}
where $C \le 3$ for eternal inflation. Note that $C=1/2$ reduces this to the earlier condition in eq. (\ref{eq1}). Similarly, the same method applies to the quantum diffusion condition, consistent with eternal inflation:
\begin{equation}\label{eternalquantumdiff}
\sqrt{8C} \cdot \frac{\Dot{\delta \phi_{\textrm{qu}}}}{|\Dot{\phi}|} =  \sqrt{2C} P_{\mathcal{R}} > 1 \quad \rightarrow \quad P_{\mathcal{R}} > \frac{1}{\sqrt{2C}}
\end{equation}
where $C=1/8$ reduces this back to quantum diffusion (\ref{eq22}) as $P_{\mathcal{R}}(k) > 2$.

Moving on to another stochastic perspective, once fluctuations exit the horizon, they cohere and behave classically. Assuming many e-foldings, these quantum fluctuations model as classical, introducing a Gaussian noise term to the Langevin equation:
\begin{equation}\label{eq9}
\Ddot{\phi} + 3 H \Dot{\phi} + \partial_{\phi}V = N(t,\Vec{x})
\end{equation}
where the right-hand side is a non-zero stochastic noise term for quantum fluctuations, including a random walk of the field $\phi$ in the potential $V(\phi)$. The evolution of the probability distribution of $\phi$ is then described by a Focker-Planck equation:
\begin{equation}\label{eq10}
\Dot{P}[\phi,t]=\frac{H^3}{8 \pi^2}\partial_i \partial^i P[\phi,t] + \frac{1}{3H} \partial_i \left(\partial^i V(\phi) P[\phi,t]\right).
\end{equation}
For a linear or quadratic potential and $H\sim$ constant, the solution becomes \cite{Rudelius:2019cfh}:
\begin{equation}\label{eq11}
P[\phi,t]=\frac{1}{\sigma(t)\sqrt{2\pi}}\exp{\left(\frac{-\left(\phi-\mu(t)\right)^2}{2 \sigma(t)^2}\right)}
\end{equation}
where $\mu(t)$ is the mean for the classical rolling $\delta \phi_{\textrm{cl}}$, and $\sigma(t)^2$ is the variance for the quantum fluctuations $\delta \phi_{\textrm{qu}}^2$.
The eternal inflation condition for this approach requires an investigation over the total volume of the universe, considering:
\begin{equation}\label{eq12}
U(\phi>\phi(\epsilon=1),t) = P_r(\phi>\phi(\epsilon=1),t) \times U(t)
\end{equation}
where $P_r \propto \int d\phi P$ is the probability of $\phi>\phi(\epsilon=1)$ decreasing with time $t$ and $U(t) \sim \exp{(3Ht)}$ is the volume of the universe increasing with time $t$. The eternal inflation condition is determined by demanding the Hubble expansion to dominate the above expression.

A simple example can be given with a linear potential $V(\phi)=V_0 - \alpha \phi$ with \cite{Rudelius:2019cfh}: $\mu(t)= \alpha t / 3H$, $\sigma(t)^2 = H^3 t / 4 \pi^2$ where the Focker-Planck equation leads in the large $t$ limit:
\begin{equation}\label{eq14}
P_r = \frac{1}{2} \textrm{erfc}\left(\frac{\frac{\alpha}{3H}t-\phi(\epsilon=1)}{\frac{H}{2\pi}\sqrt{2Ht}}\right) \sim \exp{\left(-\frac{2\pi^2\Dot{\phi}^2}{H^3}t\right)}.
\end{equation}
Then, the eternal inflation condition becomes
\begin{equation}\label{eq15}
3Ht > \frac{2\pi^2 \Dot{\phi}^2}{H^3}t \quad \rightarrow \quad \frac{2\pi^2 \Dot{\phi}^2}{3H^4} < 1.
\end{equation}
This is the same expression given in equation (\ref{eq6}), hence matches the condition $P_{\mathcal{R}} > 1/\sqrt{2C}$ with $C=3$, instead of $P_{\mathcal{R}}>1$ for $C=1/2$.

\vspace{-3mm}
\section{\label{sec:nonattractor} Non-attractor phase.}
\vspace{-4mm}
$P_{\mathcal{R}}(k)\sim \mathcal{O}(1)$ suggests a disconnection between density perturbations during eternal inflation and those responsible for observable scales' density perturbations ($P_* \simeq 2.1 \times 10^{-9}$), challenging the SR attractor assumption in inflation. Instead, a non-attractor phase emerges, where $\epsilon$ remains small while $\eta \approx \mathcal{O}(1)$ during inflation. This departure leads to post-horizon exit growth of the curvature perturbation, contrary to its customary constancy \cite{Namjoo:2012aa,Martin:2012pe,Cai:2016ngx}.

The non-attractor phase, with a constant potential $V=V_0$, yields inflationary parameters:
\begin{equation}\label{eq17}
\epsilon \propto a^{-6}, \quad \eta \simeq -6,
\end{equation}
with $\epsilon$ rapidly decaying and always large $|\eta|$. The `Graceful Exit problem' is resolved by introducing a potential $V(\phi \leq \phi_{\textrm{c}}) = V_0$ and $V(\phi > \phi_{\textrm{c}}) = V (\phi)$, with $\phi_{\textrm{c}}$ ensuring potential continuity. This phase, known as ultra slow-roll (USR)\cite{Kinney:2005vj,Tsamis:2003px}, transforms the background equation to:
\begin{equation}\label{eq18}
\Ddot{\phi} = -n H \Dot{\phi},
\end{equation}
where $n=0$ corresponds to SR, and $n=-3$ characterizes USR. Consequently, the spectral index calculation becomes \cite{Martin:2012pe}: $n_{\textrm{s}} -1 = 2 (n+3) \textrm{ for } n<-3/2.$

Given $n_{\textrm{s}} = 0.9649 \pm 0.0042$, this analysis yields $-3.01956 \lesssim n \lesssim -3.01545$, aligning with observed spectral index values. Deviation from exact USR is expected due to quantum fluctuations, so considering the Langevin equation (\ref{eq9}) instead of (\ref{eq18}) is appropriate. Incorporating this with the eternal inflation condition $P_{\mathcal{R}}>1/\sqrt{2C}$, observational bounds on e-folds (\ref{eq4}) become:
\begin{equation}\label{eq20}
\frac{\ln{\left(\sqrt{2C}P_{*}\right)}}{n_{\textrm{s,max}} -1} > \Delta N_{\textrm{min}} > \frac{\ln{\left(\sqrt{2C}P_{*}\right)}}{n_{\textrm{s,min}} -1}.
\end{equation}
Hence, for non-perpetual eternal inflation with $C=3$ and $n_{\textrm{s}} = 0.9649 \pm 0.0042$, bounds on e-folds are $485.65 \lesssim \Delta N_{\textrm{min}} \lesssim 617.65.$

\vspace{-3mm}
\section{\label{sec:stringinf}String inflation.}
\vspace{-4mm}

{ 
String theory provides a promising framework for understanding the early universe and offers a potential UV completion for inflationary dynamics. Among the various string-inspired approaches (see refs. \cite{Cicoli:2023opf,Silverstein:2016ggb,Baumann:2014nda,Burgess:2013sla,Yamaguchi:2011kg,Burgess:2011fa,Copeland:2011dx,Cicoli:2011zz} for recent reviews), we focus here on models driven by the dynamics of D-branes (and anti-D-branes), particularly in scenarios where inflation is governed by the motion of D-branes within a higher-dimensional, compactified space. These models often involve D3-branes traversing warped throats in extra dimensions, where their interactions—such as approaching, colliding, separating, or even annihilating—can trigger and sustain an inflationary phase. The two models most relevant to our discussion are:

\textbf{\textrm{i}) Inflection Point Inflation:} } Trajectories allowing for a substantial number ($\gg 60$) of e-folds of inflation can be effectively described as single-field SR scenarios \cite{McAllister:2012am}. This description becomes viable due to the extended phase of inflation occurring before the observable scales in the CMB exit the horizon, which effectively dampens non-SR and multi-field effects. Models with many e-folds exhibit phenomenology akin to inflection point inflation, a well-known representative of classical D-brane inflation cosmology. Moreover, predictions from eternal inflation patches for inflection point inflation align with the quantum cosmology of D-brane inflation \cite{Hertog:2015zwh}.

{
Considering the Kuperstein embedding \cite{Kuperstein:2004hy} for inflation near an inflection point ($\phi = \phi_0$), the phenomenological form of the potential is given with \cite{Baumann:2007np,Baumann:2007ah}
\begin{equation}
    V \simeq V_0 \left( 1 + \lambda_1 (\phi -\phi_0) + \lambda_3 (\phi - \phi_0)^3 + \dots \right),
\end{equation}
which can be well-approximated by the cubic form for some constant parameters $V_0$, $\lambda_1$ and $\lambda_3$. The total number of e-folds starting from an arbitrary point of inflaton value to the end of inflation is given by:
\begin{equation}\label{ipiTotalefolds}
N(\phi)=\left.\frac{\Delta N_{\textrm{max}}}{\pi} \cdot \arctan \left(\frac{\sqrt{\lambda_3} (\phi - \phi_0)}{\sqrt{2 \lambda_1}}\right)\right|_{\phi_{\text {end }}} ^\phi,
\end{equation}
where
$\Delta N_{\textrm{max}}$ is the maximum value of the total number of e-folds between the asymptotic limits, defined as
\begin{equation}
    \Delta N_{\textrm{max}} \equiv \pi \cdot \sqrt{\frac{2 V_{0}^2}{\lambda_1 \lambda_3}}. 
\end{equation}
This leads to the condition $\Delta N_{\textrm{max}} \geq N(\phi) \equiv \Delta N_{\textrm{min}}$ for a non-perpetual eternal inflation phase. In the limit $\lambda_3 \gg \lambda_1$, (\ref{ipiTotalefolds}) approximates to $N(\phi) \approx \Delta N_{\textrm{max}}$ when $\phi_{\textrm{end}}<\phi_0 < \phi$, indicating $\Delta N_{\textrm{max}} \simeq \Delta N_{\textrm{min}}$. This establishes a phenomenological link between D-brane inflation cosmology and non-perpetual eternal inflation.

\vspace{2mm}
\textbf{\textrm{ii}) Dirac-Born-Infeld (DBI) Inflation:} Unlike the traditional slow-roll approach, DBI inflation achieves accelerated expansion through the relativistic dynamics of D-branes moving through a warped throat in extra-dimensional space \cite{Silverstein:2003hf,Alishahiha:2004eh}. This model is characterized by non-canonical kinetic terms, which reflect the brane's motion in a higher-dimensional background. The DBI Lagrangian for a D3-brane in this context is given by:
\begin{equation}
\mathcal{L}_{\mathrm{DBI}}=--T(\phi)\left(\sqrt{1+\frac{\left(\partial_\mu \phi \partial^\mu \phi\right)}{T(\phi)}}-1\right)-V(\phi),
\end{equation}
where $T(\phi)$ is the warped brane tension and $\phi$ is the D-brane position modulus. The specifics of the brane tension and potential play a crucial role in shaping the inflationary dynamics, with the warped geometry effectively slowing the inflaton's motion.

In a non-canonical theory like DBI inflation, the speed of sound $c_s$ is not necessarily equal to the speed of light, as it would be in canonical theories. Instead, it is given by:
\begin{equation}
c_s=\left( 1+\frac{\dot{\phi}^2}{T(\phi)} \right)^{-\frac{1}{2}}.
\end{equation}
Consequently, the propagation speed of small perturbations around this homogeneous background can be slower than the speed of light. This reduced speed of sound $c_s$ modifies the standard dispersion relation for perturbations in the inflaton field, with the frequency of a wave mode now related to its wavenumber by $w^2 = c_{s}^{2} k^2$. As a result, the power spectrum of primordial perturbations is affected, leading to the expression:
\begin{equation}
   P_{\mathcal{R}}(k) =  \frac{H^2}{2 \pi \Dot{\phi}\sqrt{c_s}}.
\end{equation}
In this context, the eternal inflation condition $\delta \phi_{\textrm{qu}} > \delta \phi_{\textrm{cl}}$, as given in eq. (\ref{eq1}), can be reformulated as:
\begin{equation}
   \frac{\delta \phi_{\textrm{qu}}}{\delta \phi_{\textrm{cl}}} = \sqrt{c_s} P_{\mathcal{R}}(k) > 1 \quad \rightarrow \quad P_{\mathcal{R}} > \frac{1}{\sqrt{c_s}},
\end{equation}
where the parameterization term $\sqrt{2C}$ is directly linked to the relativistic dynamics of spacetime-filling D-branes, represented as $\sqrt{c_s}$. Consequently, the total number of e-folds for a non-perpetual eternal inflation phase, as given in eq. (\ref{eq3}), becomes:
\begin{equation}
\frac{\ln{\left(\sqrt{c_s}P_{*}\right)}}{n_{\textrm{s,max}} -1} > \Delta N_{\textrm{min}} > \frac{\ln{\left(\sqrt{c_s}P_{*}\right)}}{n_{\textrm{s,min}} -1}.
\end{equation}

Additionally, the non-canonical kinetic term in DBI inflation can lead to significant non-Gaussianities in the primordial perturbations, particularly peaking in equilateral triangle configurations \cite{Alishahiha:2004eh,Chen:2006xjb,Chen:2006nt,Chen:2010xka,Senatore:2009gt}. The level of equilateral non-Gaussianity is often parameterized by the factor $f_{\mathrm{NL}}$, which in DBI inflation is related to the sound speed $c_s$ as \cite{Alishahiha:2004eh}:
\begin{equation}
    f_{\mathrm{NL}}^{\textrm {equil}} \approx -0.32 \cdot \left(\frac{1}{c_s^2} -1\right)
\end{equation}
Thus, a smaller sound speed $c_s$ can result in large non-Gaussianities, which are constrained by observations of the CMB. The latest data \cite{Planck:2019kim} gives: $f_{\mathrm{NL}}^{\textrm {equil}}=-26 \pm 47$ (68\% C.L.), leading to a lower bound on the sound speed of $c_s \gtrsim 0.066$. Considering this lower limit, the minimum value for the total number of e-folds in a non-perpetual eternal inflation phase within the DBI inflation framework is:
\begin{equation}
    543.19 \lesssim \Delta N_{\textrm{min}} \lesssim 690.59.
\end{equation}

Building on the observations from the two string inflation models discussed, we will now analyze a toy model under the simplifying assumptions of slow-roll conditions and $c_s = 1$. This analysis will provide a quantitative demonstration of how observational predictions can align with the region of non-perpetual eternal inflation.

\vspace{-3mm}
\section{\label{sec:toymodel}A stringy toy model.}}
\vspace{-4mm}
A simple toy model for inflection point inflation is proposed with \cite{Linde:2007jn}
\begin{equation}\label{simplepot}
V(\phi)=V_0 \left(1- \frac{\lambda_p \phi^p}{p} \right).
\end{equation}
In the vicinity of the inflection point, $\phi \sim 0$, resulting in $V_\phi = 0$ and the field remaining stationary in a classical sense. Identifying $\widetilde{\phi} \sim \delta \phi_{\textrm{qu}}$ at the point where the quantum fluctuation of the scalar field matches the classical fluctuations in a non-perpetual eternal inflation phase for the local observer, the general condition (\ref{eq8}) yields
\begin{equation}\label{phistar}
\frac{H}{2\pi}\sqrt{2C} = \frac{\Dot{\widetilde{\phi}}}{H} \quad \rightarrow \quad \widetilde{\phi} = \left(\frac{H \sqrt{2C}}{2 \pi \lambda_p}\right)^{\frac{1}{p-1}}.
\end{equation}
Here, we consider a typical time interval with $t=H^{-1}$ and use the SR approximations, $\Dot{\phi} \simeq |\partial_{\phi} V|/3H$ and $3H^2 \simeq V_{0}$.

Defining the field growth during eternal inflation until a moment $\Bar{t}$ when its average value reaches $\widetilde{\phi}$, and considering:
\begin{equation}
   \delta \phi_{\textrm{qu}} \sim \widetilde{\phi} =  \frac{H}{2\pi}\sqrt{H \Bar{t}},
\end{equation}
the e-folding number during this phase is given by
\begin{equation}\label{eq24}
e^{\Bar{N}} \sim e^{H \Bar{t}} \quad \rightarrow \quad \Bar{N} = \widetilde{\phi}^2 \frac{4\pi^2}{H^2}.
\end{equation}
For the toy model (\ref{simplepot}), e-folding number and the spectral index is given by \cite{Lyth:1998xn,Kohri:2007gq,Kinney:1995cc}:
\begin{equation}
  N(\phi)= \frac{1}{p-2} \cdot \frac{1}{\lambda_p \phi^{p-2}}, \quad  n_{\textrm{s}} = 1 -  \frac{ (p-1)}{(p-2) } \cdot \frac{2}{N}.
\end{equation}
Thus, combining with eq. (\ref{phistar}) and (\ref{eq24}), one gets these expressions in terms of e-folding number during eternal inflation phase
\begin{equation}
    \Bar{N}= 2C (p-2)^2 N^2, \quad n_{\textrm{s}}=1-(p-1) \cdot \frac{2 \sqrt{2C}}{\sqrt{\Delta N - N}}
\end{equation}
where the total e-folds are given by $\Delta N = \Bar{N} + N $ considering the observational value $N=N_{\textrm{CMB}} \simeq 60$.

Similarly, applying the same method for the (eternal) quantum diffusion condition using eq. (\ref{eternalquantumdiff}) by defining the field growth due to quantum diffusion until a moment $\Bar{t}$ when its average value reaches $\Bar{\phi}$, one gets:
\begin{equation}\label{barphi}
    \Bar{\phi} = \left(\frac{\sqrt{8C}H^2}{8 \pi^2 \lambda_p}\right)^{\frac{1}{p}}.
\end{equation}
This results in the spectral index and the number of e-folds during quantum diffusion as
\begin{equation}
   \Bar{N}=\sqrt{2C} (p-2) N, \quad n_{\textrm{s}} = 1- (p-1) \cdot \frac{2 \sqrt{2C}}{\Delta N - N}.
\end{equation}

\begin{figure}[htb]
\includegraphics[width=.7325\textwidth, height=.7325\textwidth,keepaspectratio]{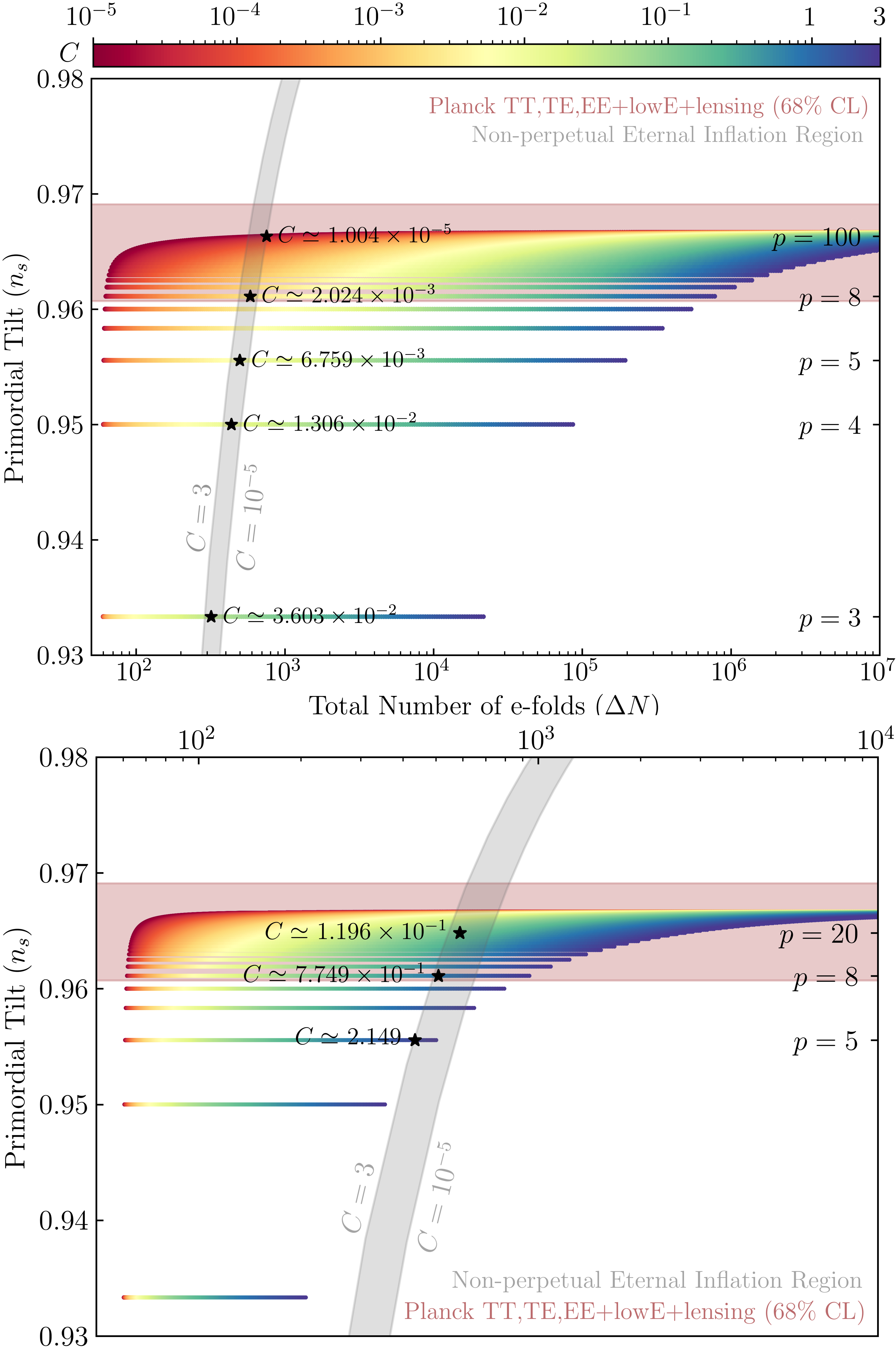}
\caption{\label{fig:1} The $C$ dependence of $(n_{\textrm{s}}$, $\Delta N)$ is illustrated for toy model inflection point inflation with selected $p$ values, considering an e-folding number $N=60$. The upper panel represents the scenario of eternal inflation, while the lower panel depicts inflation with quantum diffusion. Pink and gray shaded areas denote the model-independent allowed regions for the CMB data and non-perpetual eternal inflation, respectively. Stars corresponds to the $C$ values of the non-perpetual eternal inflation region intersecting with the model for the selected $p$ values.}
\end{figure}

In Fig. \ref{fig:1}, we observe a noteworthy consistency with a phase of non-perpetual eternal inflation in terms of $n_{\textrm{s}}$ - $\Delta N$ parameters. This consistency is pronounced for higher values of the parameter $C$ with quantum diffusion and lower values of $C$ without quantum diffusion, across various powers of the scalar field. Notably, scalar field powers exceeding $p > 20$ adhere to the quantum diffusion criterion $P_{\mathcal{R}}(k) > 2$, aligning closely with observational data on the primordial tilt. Moreover, the eternal inflation condition $P_{\mathcal{R}}(k) > 1/\sqrt{2C}$ is sufficiently satisfied for $p \geq 8$ and aligns with observations for either case.

\vspace{-3mm}
\section{\label{sec:dsconj}Emergent refined de Sitter conjecture.} 
\vspace{-4mm}
Similar to the stochastic approach, with Hubble expansion $\exp{(3Ht)}$ and the probability of inflation ending being proportional to $\Gamma \sim \exp{(3 H t)} \times \exp{\left(-t/\Bar{t}\right)}$ \cite{Barenboim:2016mmw}, the eternal inflation condition is given by
\begin{equation}\label{eq28}
\Gamma \sim \exp\left[\left(3-\frac{\delta \phi_{\textrm{qu}}^2}{\widetilde{\phi}^{2}}\right) Ht\right] \quad \rightarrow \quad \widetilde{\phi}>\frac{\delta \phi_{\textrm{qu}}}{\sqrt{3}}.
\end{equation}
For a typical quantum fluctuation $\phi_{\textrm{qu}} = H/2\pi$ and a Hilltop potential, this condition leads to \cite{Kinney:2018kew}: $\partial_{\phi}^2 V / V > -\sqrt{3}$ (in Planck units), where $\phi_{*}$ is determined by solving $P_{\mathcal{R}}(\phi_{*}) = 1$. Hence, this is equivalent to considering a non-perpetual eternal inflation phase for the local observer.

Following the same path and using eq. (\ref{eq8}) for the local view of an observer from different eternal inflation scenarios with $P_{\mathcal{R}} = 1/(\sqrt{2C})$, one can recast that inequality. Considering the toy model (\ref{simplepot}) with $p=2$ for a simple Hilltop potential, using eq. (\ref{phistar}) yields the inequality (\ref{eq28}) as
\begin{equation}
\frac{H \sqrt{2C}}{2 \pi \lambda_2} > \frac{H}{2 \pi \sqrt{3}} \quad \rightarrow \quad \sqrt{6C}>\lambda_2.
\end{equation}
In the vicinity of the inflection point, $\phi \sim 0$, one can identify $\partial_{\phi}^2 V \simeq - V_0 \lambda_2 $ or $\eta \simeq -\lambda_2$. Hence, one obtains
\begin{equation}
\frac{\partial_{\phi}^2 V}{V}>-\sqrt{6C}.
\end{equation}

Now, considering the eternal inflation condition $P_{\mathcal{R}} > 1/\sqrt{2C}$ with $C \le 3$, any curvature perturbation amplitude at or below $1/\sqrt{6}$ indicates it is not eternal. Taking the maximum limit, with $C = 3$, results in $\partial_{\phi}^2 V / V > -\sqrt{18} \sim -4.24$. This connects to the refined de Sitter conjecture \cite{Ooguri:2018wrx,Garg:2018reu}. For eternal inflation, an upper bound to the parameter $c'$ of the refined de Sitter conjecture is obtained as:
\begin{equation}\label{eq31}
-c' \ge \frac{\partial_{\phi}^2 V}{V} > -4.24 \quad \rightarrow \quad 4.24 > c',
\end{equation}
or more generally, $ \sqrt{6C} > c'$. Thus, different non-perpetual eternal inflation scenarios marginally align with the refined de Sitter Conjecture. Eternal inflation becomes a characteristic feature of the `landscape' when the specified condition is satisfied, particularly with $C$ values roughly in the range of $1/16$ to $3$, or when $c'$ is on the order of unity, i.e. $c'=\mathcal{O}(1) \simeq (0.61 \sim 4.24)$, as implied by the conjecture. Applying the same method for quantum diffusion using eq. (\ref{barphi}), one obtains a similar result with $3\sqrt{2C} > c'$. Thus, for values of $C$ around $1/32$ to $3$, one obtains $c'=\mathcal{O}(1) \simeq (0.75 \sim 7.34).$

\vspace{-3mm}
\section{Discussion.} 
\vspace{-4mm}
In this letter, we generalize the criteria for eternal inflation to encompass nearly every conceivable scenario. Introducing a bound on the amplitude of curvature perturbation, defined as $P_{\mathcal{R}} = 1/\sqrt{2C}$ where $C\leq3$, we imply a finite lifetime for eternal inflation, rendering it non-perpetual from the perspective of a local observer. Our model-independent approach establishes a minimum requirement of $\Delta N_{\textrm{min}} \approx 485$ e-foldings for eternal inflation to align with observational CMB data. It's important to note that, with the parameterizing term $C$ in play, the high curvature fluctuations across various eternal inflation cases are encapsulated within $C$. In addition, the presented view may extend beyond a simple homogeneous picture and holds relevance for different values of $C$ within the same model.

{ To ground these general criteria in specific scenarios, we examined inflection point inflation and DBI inflation, which illustrate how these generalized criteria apply across different theoretical frameworks. In inflection point inflation, the extended slow-roll regime naturally produces a large number of e-folds, aligning well with the conditions for non-perpetual eternal inflation. Meanwhile, DBI inflation introduces a reduced sound speed $c_s$,  which modifies the power spectrum and generates significant non-Gaussianities—both observable and subject to constraints from CMB data. Crucially, the parameterization term $C$  in our framework is directly linked to the relativistic dynamics of spacetime-filling D-branes through the sound speed, further illustrating how these dynamics fit within our generalized framework.}

However, our determination of this threshold assumes a mild-scale dependence of the primordial tilt, using the observational CMB value as a reference point. This assumption is further elaborated for the non-attractor phase, where an exact USR phase aligns with  $n_{\textrm{s}} = 1$ for a constant potential \cite{Martin:2012pe}. Despite the likelihood of the primordial tilt being scale-dependent in a more realistic scenario, the methodology for determining the minimum required total number of e-folds remains robust, considering a fixed maximum value as $n_{\textrm{s}}(k_*) \leq n_{\textrm{s,max}}(k_*)$, as firstly denoted in \cite{Kinney:2014jya}. Notably, this limit can be further refined for a more realistic case by considering the running of the primordial tilt and its higher derivatives to all orders \cite{WMAP:2003syu,Garcia-Bellido:2014gna,Zarei:2014bta,Munoz:2016owz,Gron:2018rtj}. This entails accounting for the scale dependence of the curvature perturbation amplitude, as described by the expression
\begin{equation}
    P_{\mathcal{R}}(k)=P_{*} \left(\frac{k}{k_*}\right)^{n_{\textrm{s}}(k_*) -1 + \alpha \ln{\left(\frac{k}{k_*}\right)} + \dots}, 
\end{equation}
and incorporating non-linear effects on superhorizon scales. Hence, this approach offers a compelling and model-independent method for testing eternal inflation predictions in a more detailed manner. Alternatively, exploring the implications of non-perturbative methods like the $\delta N$ formalism \cite{Enqvist:2008kt,Fujita:2013cna,Vennin:2015hra,Pattison:2017mbe,Pattison:2021oen} may provide a promising avenue for future research, offering deeper insights into the dynamics of eternal inflation and its observational consequences.

We then illustrate how predictions from various non-perpetual eternal inflation phases can be rigorously tested against observational data on the $(n_{\textrm{s}}, \Delta N)$ plane, utilizing a toy model of inflection point inflation. Extending our analysis to incorporate a Hilltop model, as outlined in the seminal study by Kinney \cite{Kinney:2018kew}, where the parameterization is now set as $P_{\mathcal{R}} = 1  \rightarrow  1/(\sqrt{2C})$, we establish a connection between the parameters $C$ and $c'$ in the refined de Sitter conjecture. Remarkably, even at $C=3$, $c'$ remains close to unity, suggesting its emergence from the criteria of eternal inflation. These findings imply that low-energy effective field theories in the landscape reveal an eternal nature, driven by diverse values of $c'$ or $C$. Considering the refined de Sitter conjecture in conjunction with string theory's UV completion, the emergence of a landscape-based 'multiverse' scenario may become a harmonious feature within the evolving cosmic landscape shaped by eternal inflation.
\vspace{3mm}
\begin{acknowledgments}
\textbf{Acknowledgments:} The work of OG is supported in part by the Istanbul Technical University Research Fund under grant number TGA-2024-45577.
\end{acknowledgments}

%\appendix

%\section{Appendixes}

%\nocite{*}

\bibliography{apssamp}% Produces the bibliography via BibTeX.

\end{document}